\documentstyle[preprint,aps]{revtex}
\newcommand{\beq}{\begin{equation}}
\newcommand{\eeq}{\end{equation}}
\newcommand{\beqa}{\begin{eqnarray}}
\newcommand{\eeqa}{\end{eqnarray}}
\newcommand{\ba}{\begin{array}}
\newcommand{\ea}{\end{array}}

\begin{document}
\draft

\widetext 

\title{Thermodynamics of Bose-Condensed Atomic Hydrogen} 
\author{B. Pozzi$^{(1)(2)}$, L. Salasnich$^{(1)(2)}$, 
A. Parola$^{(1)(3)}$, and L. Reatto$^{(1)(2)}$ } 
\address{
$^{(1)}$ Istituto Nazionale per la Fisica della Materia, Unit\`a di Milano,\\ 
Via Celoria 16, 20133 Milano, Italy \\ 
$^{(2)}$ Dipartimento di Fisica, Universit\`a di Milano, \\ 
Via Celoria 16, 20133 Milano, Italy \\ 
$^{(3)}$ Dipartimento di Scienze Fisiche, Universit\`a dell'Insubria, \\ 
Via Lucini 3, 23100 Como, Italy} 

\maketitle

\begin{abstract} 
We study the thermodynamics of the Bose-condensed 
atomic hydrogen confined in the Ioffe-Pritchard potential. 
Such a trapping potential, that models the magnetic trap used 
in recent experiments with hydrogen, is anharmonic and strongly 
anisotropic. We calculate the ground-state properties, 
the condensed and non-condensed fraction and 
the Bose-Einstein transition temperature. 
The thermodynamics of the system is strongly affected by the 
anharmonicity of this external trap. 
Finally, we consider the possibility to detect Josephson-like 
currents by creating a double-well barrier with a laser beam.  
\end{abstract}
\vskip 1. truecm
\pacs{PACS numbers: 03.75.Fi, 05.30.Jp, 32.80.Pj}

\narrowtext

Few years ago, Bose-Einstein condensation (BEC) 
has been experimentally observed in clouds of trapped 
alkali-metal atoms [1]. Recently, BEC has been also achieved 
with atomic hydrogen confined in a Ioffe-Pritchard trap [2]. 
That is an important 
result because hydrogen properties, like interatomic 
potentials and spin relaxation rates, 
are well understood theoretically. 
As stressed by Killian {\it et al.} [2], 
the s-wave scattering length of the hydrogen 
is very low and, compared with other atomic species, 
the condensate density is high, even for small condensate fractions. 
Moreover, due to hydrogen's small mass, the BEC transition occurs 
at higher temperatures than in those of alkali atoms. 

In this paper we calculate the thermodynamical properties 
of the trapped hydrogen gas by using the quasi-classical 
Hartree-Fock approximation [3]. This approach is justified 
by the very large number of atoms (about $10^{10}$) in the 
trap and by the relatively high temperatures involved (order of $\mu$K). 
Due to the anharmonic external trap, 
the analytical results for BEC thermodynamics 
obtained by Stringari {\it et. al} [3] cannot be used for quantitative 
predictions. Our detailed theoretical study of the hydrogen 
thermodynamics can give useful informations for future experiments 
with a better optical resolution. 
In the last part of the paper we discuss the criteria 
for macroscopic quantum tunneling and macroscopic 
quantum self-trapping by using a laser beam to create 
a double-well potential. 

In the experiment reported in Ref. [2], the axially symmetric 
magnetic trap is modelled by the Ioffe-Pritchard potential 
\beq 
U(\rho , z) = \sqrt{(\alpha \rho)^2 + 
(\beta z^2+\gamma)^2} - \gamma  \; ,
\eeq 
where $\rho$ and $z$ are cylindrical coordinates and 
the parameters $\alpha$, $\beta$ and $\gamma$ 
can be calculated from the magnetic coil geometry. 
In particular, for small displacements, 
the radial oscillation frequency 
is $\omega_{\rho}=\alpha /\sqrt{m\gamma}=2\pi \times 3.90$ kHz, 
the axial frequency is $\omega_z=\sqrt{2\beta /m}=
2\pi \times 10.2$ Hz and $\gamma/k_B=35$ $\mu$K [2]. 

A dilute gas of $N$ identical hydrogen atoms is theoretically 
described by a bosonic field operator ${\hat \psi}({\bf r},t)$. 
Following a standard treatment, 
one can separate out the condensate part with the Bogoliubov 
prescription ${\hat \psi}({\bf r},t)=\Phi({\bf r})+{\hat \phi}({\bf r},t)$, 
where $\Phi({\bf r}) =\langle {\hat \psi}({\bf r},t)\rangle$ 
is the macroscopic wavefunction (order parameter) 
of the condensate, normalized to the number $N_0$ of condensed atoms, 
and $\langle ... \rangle$ is the mean value in the 
grand-canonical ensemble. 
Then, in the mean-field approximation, the order parameter $\Phi({\bf r})$ 
satisfies the following finite-temperature Gross-Pitaevskii  
equation [4,5] 
\beq  
\left[ -{\hbar^2 \over 2 m} \nabla^2 + U({\bf r}) 
+ g n_0({\bf r}) + 2 g n_T({\bf r})
\right] \Phi({\bf r}) = \mu \Phi({\bf r})   \; ,  
\eeq 
where $m$ is the mass of the atomic hydrogen, 
${\mu}$ the chemical potential, and $g={4\pi \hbar^2 a/m}$ 
is the scattering amplitude with $a$ the s-wave scattering 
length ($a=0.0648$ nm). 
The condensate density is $n_0({\bf r})=|\Phi({\bf r})|^2$ 
and $n_T({\bf r})=<{\hat \phi}^+({\bf r},t){\hat \phi}({\bf r},t)>$ 
is the thermal density of non-condensed particles, 
normalized to $N_T=N-N_0$. 
Notice that we ignore the $T=0$ quantum depletion. 
In fact, the corrections 
to the mean-field results are fixed by the gas parameter 
$n({\bf 0}) a^3$, where the total density $n({\bf r})$ is 
evaluated at the center of the trap [6]. 
We shall show that this parameter is very small 
also when there are billions of atoms in the trap. 

The thermal density can be calculated through 
the quasi-classical Hartree-Fock approximation 
\beq
n_T({\bf r})={1\over \lambda^3}
\; g_{3/2}\left( e^{-(U({\bf r})+2g n({\bf r}) -\mu )/k_BT}\right) \; ,
\eeq
where $g_{\alpha}(z)=\sum_{k=1}^{\infty}z^k/k^{\alpha}$ 
and $\lambda=(2\pi \hbar^2 /m k_B T)^{1/2}$ is the thermal length. 
Thus, the thermal particles behave as "non-interacting" bosons 
moving in the self-consistent effective potential $U({\bf r})+ 
2 g n({\bf r})$, where the term $2 g n({\bf r})$ is the mean-field 
generated by interactions with other atoms [3]. 
Note that the quasi-classical approximation is accurate 
for the experimental hydrogen cloud, which has $N=10^{10}$ atoms. 
Also the Hartree-Fock approximation is valid in this context because 
$k_B T$ is much larger than the chemical potential $\mu$. 

In many papers, the thermodynamics of Bose-condensed dilute gases 
has been studied by solving the Eq. (2) and (3) with a 
self-consistent iterative procedure (see [6] and references 
therein). We shall adopt this 
method that, in our case, describes correctly both the 
high- and low-temperature regimes due to the negligible effect 
of collective excitations in the thermal density. 
It is important to observe that mean-field predictions  
for trapped alkali-metal atoms are in good agreement 
with recent path-integral Monte Carlo calculations [7]. 
Moreover, in the experiments with atomic hydrogen, the gas is dilute 
($n a^3\ll 1$) but also strongly interacting because 
$N a/a_H \gg 1$, where 
$a_H=(\hbar /m \omega_H)^{1/2}$ with $\omega_H=
(\omega_{\rho}^2\omega_z)^{1/3}$. It follows that the kinetic 
term of the Eq. (2) can be safely neglected, as 
we have verified by numerically solving Eq. (2) at zero temperature, 
and one gets the Thomas-Fermi condensate density 
\beq 
n_0({\bf r},t) = {1\over g} 
\left[\mu -  U({\bf r}) - 2 g n_T({\bf r}) \right] \; ,
\eeq
in the region where $\mu > U({\bf r}) + 
2 g n_T({\bf r})$, and $n_0({\bf r})=0$ outside. 
In practice, we study the BEC thermodynamics by solving  
self-consistently the Eq. (3) and (4). 

First, let us consider a noninteracting gas ($g=0$). 
In such a case, the formula (3) of 
thermal density simplifies and one 
can directly obtain the BEC transition 
temperature $T_c^0$ as a function of 
the total number $N$ of atoms by numerically 
solving the equation 
\beq
N= \int { d^3{\bf r} \over (2\pi \hbar^2 /m k_B T_c^0)^{3/2}} 
\; g_{3/2}\left( e^{-U({\bf r})/k_B T_c^0}\right) \; . 
\eeq 
In Fig. 1 we compare the BEC transition temperature $T_c^0$ 
of the Ioffe-Pritchard potential with the analytic formula 
$T_c^{0,H}=0.94 \hbar \omega_H N^{1/3}/k_B$, 
that is exact in the thermodynamic limit for 
the harmonic potential $U=(m/2)(\omega_\rho^2 \rho^2+ 
\omega_z^2 z^2)$. Fig. 1 shows that, for a large number of atoms, 
$T_c^{0,H}$ exceeds $T_c^0$. In particular, 
for $N=2\cdot 10^{10}$ the relative difference is more 
than $41\%$. In the Ioffe-Pritchard trap the 
BEC transition temperature is strongly affected by the 
fact that along the cylindrical radius the quadratic behavior 
becomes almost linear at large distances 
(for a discussion of potentials with exact power laws see [8]). 
It means that the density of states is higher than in the harmonic case 
and consequently $T_c^0$ is suppressed. 
The critical temperature in this Ioffe-Pritchard trap 
can be well represented by the law $T_c^0=b N^{1/\eta}$, 
where $b=5.47\cdot 10^{-2}$ $\mu$K and $\eta=3.51$. 

The role of the interatomic interaction 
on the transition temperature $T_c$ is very small. 
In Fig. 1 one observes that the repulsive interaction reduces $T_c$ 
both in harmonic and Ioffe-Pritchard traps. The shift of the 
transition temperature $\delta T_c$ is in agreement 
with the law $\delta T_c/T_c = - 1.3 (a/a_H) N^{1/6}$, predicted by 
Stringari {\it et al.} [3] for the harmonic trap. 
These corrections are of the order of $0.1$ $\mu$K. 

In Fig. 2 we show the condensate density 
$n_0({\bf r})$ and the thermal density $n_T({\bf r})$ 
at $T=45$ and $47$ $\mu$K for $N=2.2\cdot 10^{10}$ atoms.  
The thermal density shows a depletion near the origin, that is 
due to the presence of the condensate fraction in that region. 
Note that the thermal cloud fills a very large spatial region 
compared to the condensate one. 

In Fig. 3 we present various properties of the condensate  
as a function of temperature for $N=2.2\cdot 10^{10}$ atoms. 
We consider the number of atoms and temperatures achieved 
in the MIT experiment [2]. The central density and the size 
of the condensate are particularly interesting 
because can be easily detected experimentally. 
Our results are compatible with the experimental data: 
at $T=45\pm5$ $\mu$K with $N=2.2 \cdot 10^{10}$ the estimated 
condensate fraction is $5\%$, the peak 
condensate density $4.8\pm 1.1\cdot 10^{15}$ cm$^{-3}$, 
the condensate diameter is $d=15$ $\mu$m and its length $l=5$ mm [2]. 
We note that, as previously anticipated, the gas parameter 
$n({\bf 0})a^3$ is always less than $10^{-6}$. 

The condensed fraction, the energy per particle of the condensate 
and the chemical potential are shown in Table 1. We calculate also 
the two-body decay rate 
$\Gamma=c\int d^3{\bf r} n_0^2({\bf r})$, where 
$c=1.1\cdot 10^{-15}$ cm$^3$/s [3]. The decay time can be estimated 
as $\tau_{1/2}=7/(2cn_0({\bf 0}))$ [3]. 
The experimentally measured life-time (about $5$ seconds 
with $N_0/N =0.05$) is larger than $\tau_{1/2}$. 
As stressed in [2], the thermal gas continually replenishes 
the condensate as atoms are lost through two-body collisions. 
Thus the apparent life-time of the condensate is increased 
by this replenishment. 

For the sake of completeness, in Fig. 4 we plot 
the condensate fraction $N_0/N$ as a function of temperature 
for $N=10^8$, $10^9$ and $10^{10}$ atoms. 
Note that Hijmans, Kagan, Shlyapnikov and Walraven [9] 
have shown that, due to the balance between the thermalization 
rate and the two-body spin-relaxation decay rate,  
the maximum achievable equilibrium condensate fraction for 
hydrogen cannot be very large. 
Moreover, a recent theoretical paper [10] suggests the 
possibility of interesting nonequilibrium effects 
like the short-time formation of quasicondensate droplets. 
However, we belive that the equilibrium results 
we have shown may provide a useful guide for future experiments 
with a better optical resolution. 

A very interesting issue is the possibility to detect macroscopic 
quantum tunneling (MQT) and Josephson-like oscillations of the 
Bose condensate in double-well traps [11,12]. Recently, we have shown that 
with $^{23}$Na atoms in harmonic trap one sees only 
the macroscopic quantum self-trapping (MQST) 
of the condensate. To get outside the MQST regime 
it is necessary to strongly reduce the scattering 
length [12]. Due to its very low scattering length, 
atomic hydrogen is a good candidate for MQT in double-well traps. 

It is easy to create a double-well trap for a cigar shaped 
condensate by using a laser beam [13]. The effect of a laser beam 
on atoms can be modelled by the following potential 
\beq
U_{L}(z)=U_0 \exp{ \Big( {-z^2\over 2 \sigma^2} \Big)} \; , 
\eeq 
where the potential barrier $U_0$ is proportional 
to the total power of the 
laser beam perpendicular to the long axis of the condensate, and 
$\sigma$ is the beam radius [13]. 

Smerzi {et al.} [11] have shown that the time-dependent 
behavior of the condensate in a double-well potential 
depends on the parameter $\Lambda = 4 E_{int}/|\Delta E^0|$: 
$E_{int}$ is the interaction energy of the condensate 
and $\Delta E^0$ is the kinetic+potential energy 
splitting between the ground state and the quasi-degenerate 
odd first excited state of the GP equation [12,13]. 
Let $z=(N_1-N_2)/N$ be the fractional population 
imbalance of the condensate in the two wells. 
For a fixed $\Lambda$ ($\Lambda >2$) 
there exists a critical $z_c =2\sqrt{\Lambda -1}/\Lambda$ 
such that for $0 < z < z_c$ there are Josephson-like 
oscillations of the condensate. For $z\ll z_c$ these oscillations are 
harmonic with period 
$\tau = \tau_0/\sqrt{1+\Lambda}$, where 
$\tau_0 =2\pi \hbar /|\Delta E^0|$. 
Instead for $z_c<z \leq 1$ there is MQST of the condensate: 
even if the populations 
of the two wells are initially set in an asymmetric state ($z\ne 0$)
they maintain the original population imbalance without transferring
particles through the barrier [11,12]. 

By solving the zero-temperature GP equation (2) with the 
steepest-descent method in the double-well trap 
given by Eq. (1) and Eq. (6), 
we calculate the parameter $\Lambda$ for the condensate of hydrogen 
atoms in the tunneling region. 
In Table 2 it is shown that, for a small number of 
condensed atoms (about $10^4$), the MQT is possible 
with a relatively large fractional population imbalance $z$. 
For this number of condensed atoms, 
the period $\tau$ of oscillation is smaller 
than the life-time $\tau_{1/2}$ of the condensate, 
that is about $50$ seconds. 
For a larger number of atoms, the parameter $\Lambda$ quickly grows 
and the condensate remains self-trapped. 
Note that at nonzero temperature, BEC depletion and 
thermal fluctuations will slightly modify the 
parameters of the tunneling and will 
damp the coherent oscillations. 
Nevertheless, as shown by Zapata, Sols and Leggett [14], 
the effect of damping is negligible 
for temperatures lower than about $10 \hbar \omega_H$. 

In this paper we have studied the thermodynamics 
of the atomic hydrogen in a Ioffe-Pritchard trap. 
We have found that the BEC transition temperature is 
strongly dependent on the anharmonicity of the 
trap while the effect of repulsive interaction is small. 
We have calculated the properties of the Bose condensate 
at various temperatures. Our results are consistent with available 
experimental data, which still have a large error. 
The present paper should be very useful 
for future experiments because it gives detailed information 
on the Bose condensate at temperatures not achieved so far. 
Finally, we have considered the inclusion of a laser beam 
to produce Josephson-like currents in the resulting double-well trap. 
Our calculations suggest that with a small number of condensed 
atoms (about $10^{4}$) the macroscopic quantum tunneling can  
be observed. 

\vskip 1. truecm
\section*{ACKNOWLEDGEMENTS}
\par
This work has been supported by the INFM Research Advanced Project 
on Bose-Einstein Condensation. 

\newpage
\section*{References}
\vskip 1. truecm

\begin{description}

\item{\ 1.} M.H. Anderson {\it et al.}, Science {\bf 269}, 189 (1995); 
K.B. Davis {\it et al.}, Phys. Rev. Lett. {\bf 75}, 
3969 (1995); C.C. Bradley {\it et al.}, 
Phys. Rev. Lett. {\bf 75}, 1687 (1995). 

\item{\ 2.} T.C. Killian {\it et al.}, Phys. Rev. Lett. 
{\bf 81}, 3807 (1998); D.G. Fried {\it et al.}, 
Phys. Rev. Lett. {\bf 81}, 3811 (1998); 
T.J. Greytak {\it et al.}, submitted to Physica B. 
 
\item{\ 3.} S. Giorgini, L.P. Pitaevskii, and S. Stringari, 
Phys. Rev. A {\bf 54}, 4633 (1996); Phys. Rev. Lett. {\bf 78}, 3987 (1997); 
J. Low Temp. Phys. {\bf 109}, 309 (1997). 

\item{\ 4.} A. Griffin, Phys. Rev. B {\bf 53}, 9341 (1996). 

\item{\ 5.} E.P. Gross, Nuovo Cimento {\bf 20}, 454 (1961); 
J. Math. Phys. {\bf 4}, 195 (1963); 
L.P. Pitaevskii, Zh. Eksp. Teor. Fiz. {\bf 40}, 646 (1961) 
[Sov. Phys. JETP {\bf 13}, 451 (1961)]. 

\item{\ 6.} F. Dalfovo, S. Giorgini, L.P. Pitaevskii, and S. 
Stringari, Rev. Mod. Phys. {\bf 71}, 463 (1999). 

\item{\ 7.} W. Krauth, Phys. Rev. Lett. {\bf 77}, 3695 (1996); 
M. Holzmann, W. Krauth, and M. Naraschewski, 
Phys. Rev. A {\bf 59}, 2956 (1999). 

\item{\ 8.} V. Bagnato, D.E. Pritchard and D. Kleppner, 
Phys. Rev. A {\bf 35}, 4354 (1987); P.W.H. Pinkse {\it et al.}, 
Phys. Rev. Lett. {\bf 78}, 990 (1997). 

\item{\ 9.} T.W. Hijmans, Yu. Kagan, G.V. Shlyapnikov and 
J.T.M. Walraven, Phys. Rev. B {\bf 48}, 12886 (1993). 

\item{\ 10.} R. Cot\`e and V. Kharchenko, Phys. Rev. Lett. 
{\bf 83}, 2100 (1999). 

\item{\ 11.} A. Smerzi, S. Fantoni, S. Giovanazzi, and S.R. Shenoy, 
Phys. Rev. Lett. {\bf 79}, 4950 (1997); S. Raghavan, A. Smerzi, 
S. Fantoni, and S.R. Shenoy, Phys. Rev. A {\bf 59}, 620 (1999). 

\item{\ 12.} L. Salasnich, A. Parola, and L. Reatto, Phys. Rev. A 
{\bf 60}, 4171 (1999). 

\item{\ 13.} M.R. Andrews {\it et al.}, Science {\bf 275}, 637 (1997). 

\item{\ 14.} I. Zapata, F. Sols and A.J. Leggett, 
Phys. Rev. A {\bf 57}, R28 (1998). 

\end{description}
\newpage

\begin{center}
\begin{tabular}{|cccccc|} 
\hline \hline 
$T$ ($\mu$K) & $\;\;\;\;\; N_0/ N \;\;\;\;\;$ & $E/ N_0$ & 
$\;\;\;\;\;\; \mu \;\;\;\;\;\;$  & $\;\;\;\;\; \Gamma \;\;\;\;\; $ 
& $ \;\;\;\; \tau_{1/2} \;\;\;\;$  \\ 
\hline 
$44.0$  & $0.21$  & $4.64$ & $6.59$ & $2.53$ & $0.37$  \\
$44.5$  & $0.18$  & $4.39$ & $6.25$ & $2.08$ & $0.38$  \\
$45.0$  & $0.16$  & $4.13$ & $5.89$ & $1.65$ & $0.41$  \\ 
$45.5$  & $0.13$  & $3.82$ & $5.46$ & $1.27$ & $0.44$  \\
$46.0$  & $0.10$  & $3.45$ & $4.95$ & $0.88$ & $0.49$  \\
$46.5$  & $0.07$  & $3.04$ & $4.37$ & $0.56$ & $0.56$  \\
$47.0$  & $0.04$  & $2.52$ & $3.64$ & $0.29$ & $0.69$  \\
$47.5$  & $0.02$  & $1.89$ & $2.75$ & $0.10$ & $0.93$  \\
\hline \hline 
\end{tabular} 
\end{center} 
{\small TAB. 1. Condensate properties, for 
$N=2.2\cdot 10^{10}$ interacting hydrogen atoms in the 
Ioffe-Pritchard trap, at various temperatures: 
condensed fraction $N_0/N$, energy per particle of the condensate 
$E/N_0$ and chemical potential $\mu$ in units of $10^3 \hbar \omega_z$ 
(where $10^3 \hbar \omega_z/k_B=0.49$ $\mu$K), 
two-body decay rate $\Gamma$ in units of $10^{10}$ s$^{-1}$ 
and decay time $\tau_{1/2}$ in seconds.}

\vskip 1.0cm 

\begin{center}
\begin{tabular} {|c|ccccccc|}
\hline \hline 
$\;\;N\;\;$  & $\;\;U_0\;\;$  & $\;\;E_{int}\;\;$  & $\;\;\Delta E^0\;\;$ & 
$\;\;\Lambda\;\;$  & $\;\;\; z_c \;\;\;$ & $\;\;\; z_c N \;\;\;$ 
& $\;\;\; \tau \;\;\;$ \\ 
\hline
$10000$  & 8   & 1.371  & 0.222 & 24.69  & 0.39 & $3900$  & $0.09$   \\
$10000$  & 10  & 1.383  & 0.164 & 33.64  & 0.34 & $3400$  & $0.10$   \\
$10000$  & 12  & 1.397  & 0.126 & 44.38  & 0.30 & $3000$  & $0.12$   \\
\hline
$50000$  & 16  & 4.345  & 0.095 & 183.03 & 0.15 & $7500$  & $0.08$   \\
$50000$  & 18  & 4.361  & 0.072 & 242.95 & 0.13 & $6500$  & $0.09$   \\
$50000$  & 20  & 4.378  & 0.055 & 317.80 & 0.11 & $5500$  & $0.10$   \\
\hline \hline
\end{tabular}
\end{center}
{\small TAB. 2. Parameters of the MQT for $N=10^{4}$ 
and $N=5\cdot 10^4$ condensed hydrogen atoms. 
Ioffe-Pritchard external potential and different 
values of the double-well potential barrier $U_0$. 
The laser-beam radius is $\sigma = 5$ $\mu$m. 
The energies are in units of $\hbar \omega_z$ and 
the oscillation period $\tau$ in seconds.}  
\newpage

\section*{Figure Captions}

{\bf Fig. 1}. BEC transition temperature $T_c$
versus number $N$ of hydrogen atoms. Comparison between
harmonic and Ioffe-Pritchard trapping potential.
\vskip 1. truecm

{\bf Fig. 2}. Condensate density profile
$n_0$ and thermal density profile $n_T$,
for a cloud of $N=2.2\cdot 10^{10}$ interacting hydrogen atoms
in the Ioffe-Pritchard potential. Densities are in units of $a_z^3$.
Notice the very different abscissae scales.
\vskip 1. truecm

{\bf Fig. 3}. Observables of the Bose condensate as a function
of temperature $T$. $N=2.2\cdot 10^{10}$ interacting hydrogen atoms
in the Ioffe-Pritchard trap.
\vskip 1. truecm

{\bf Fig. 4}. Condensate fraction $N_0/N$ as a function
of temperature $T$. $N$ interacting hydrogen atoms
in the Ioffe-Pritchard trap.

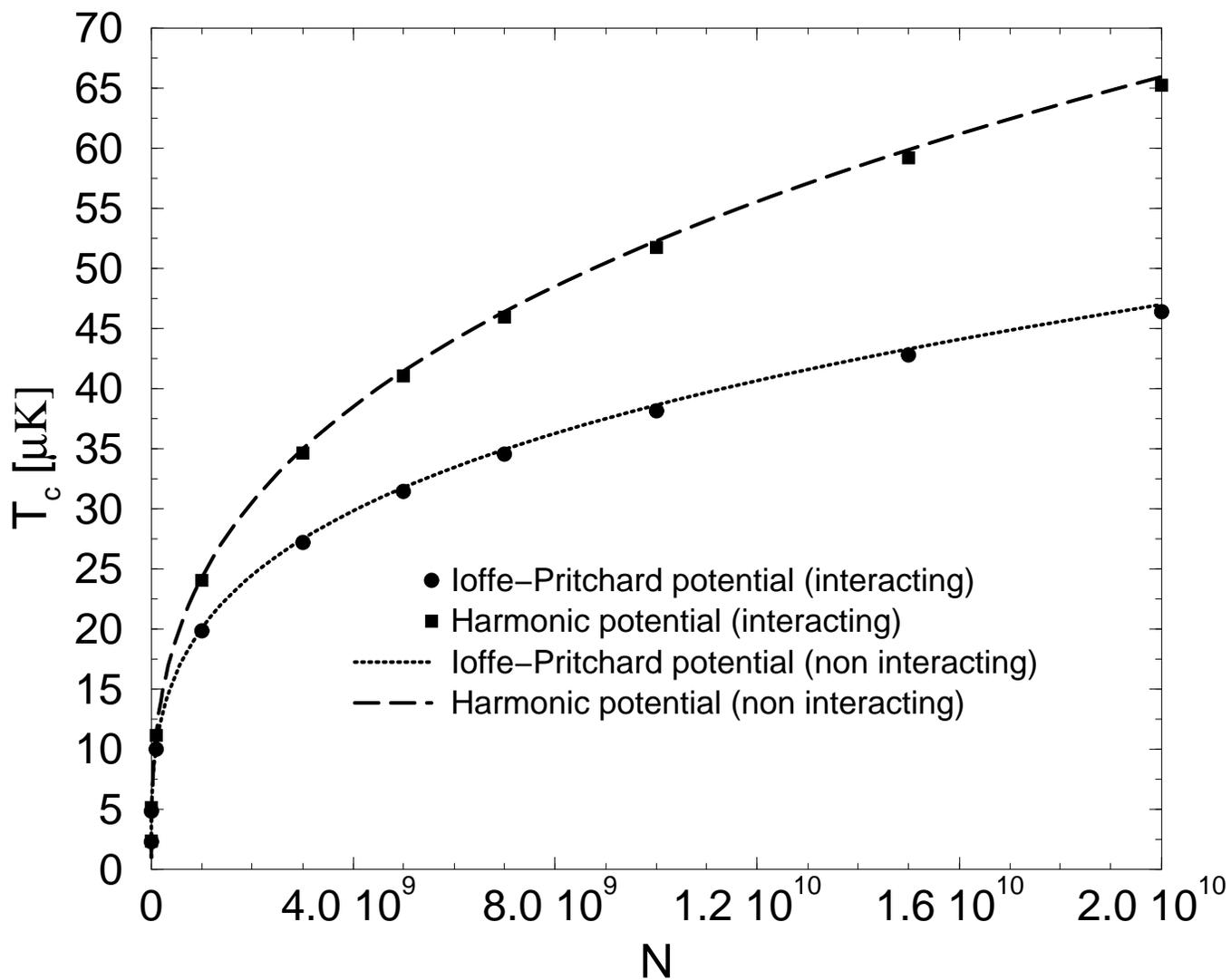
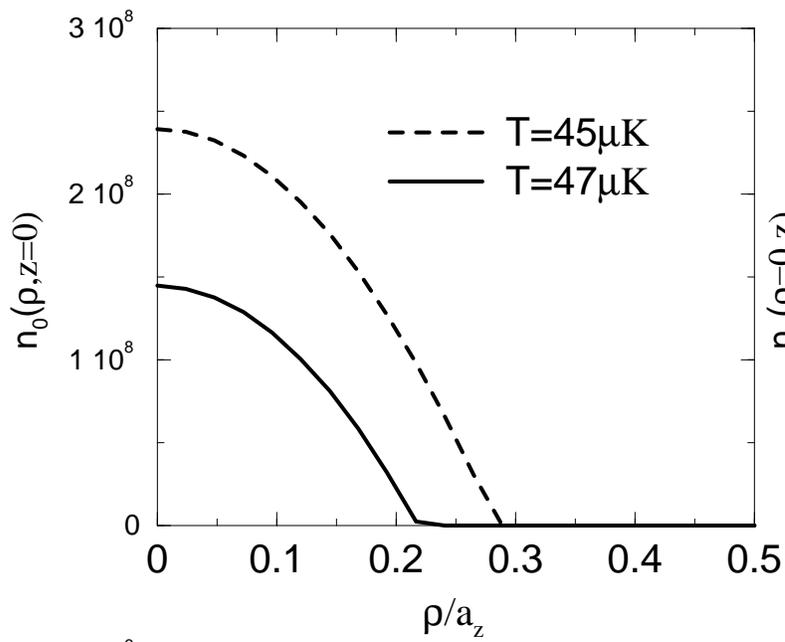
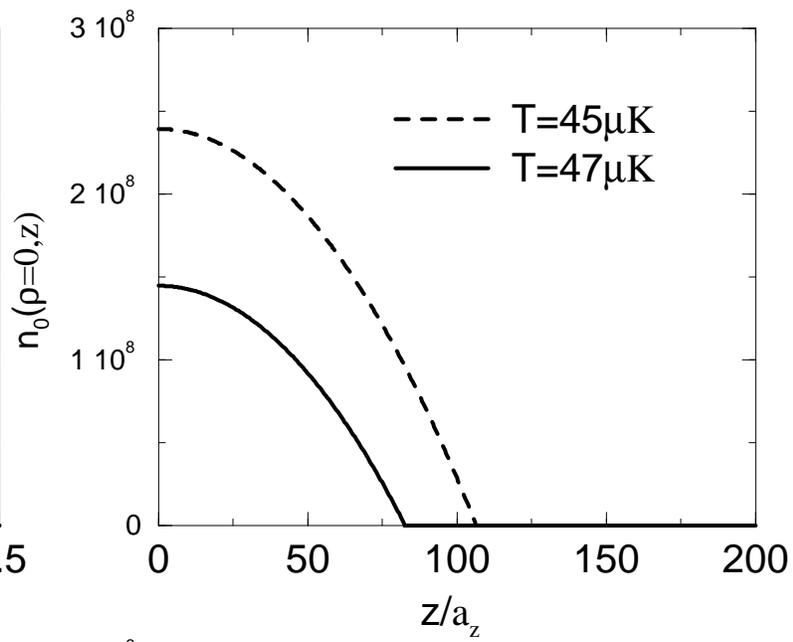
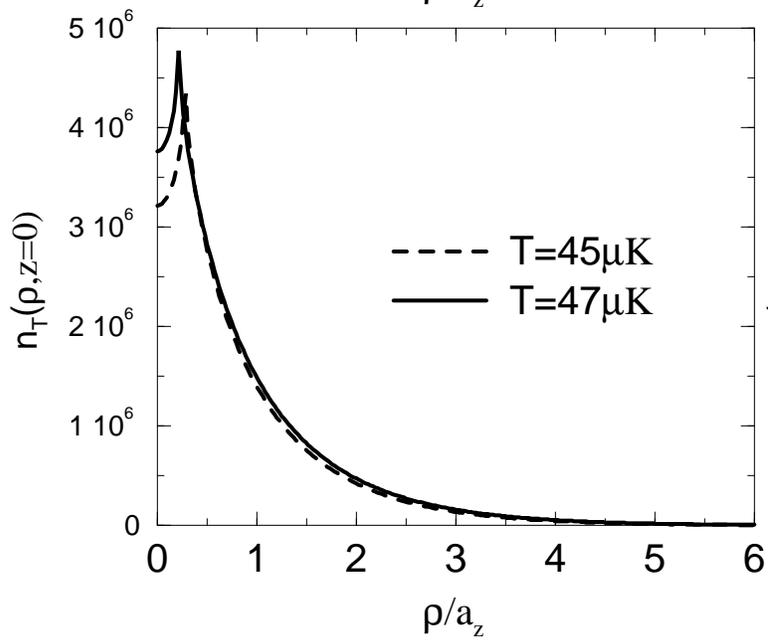
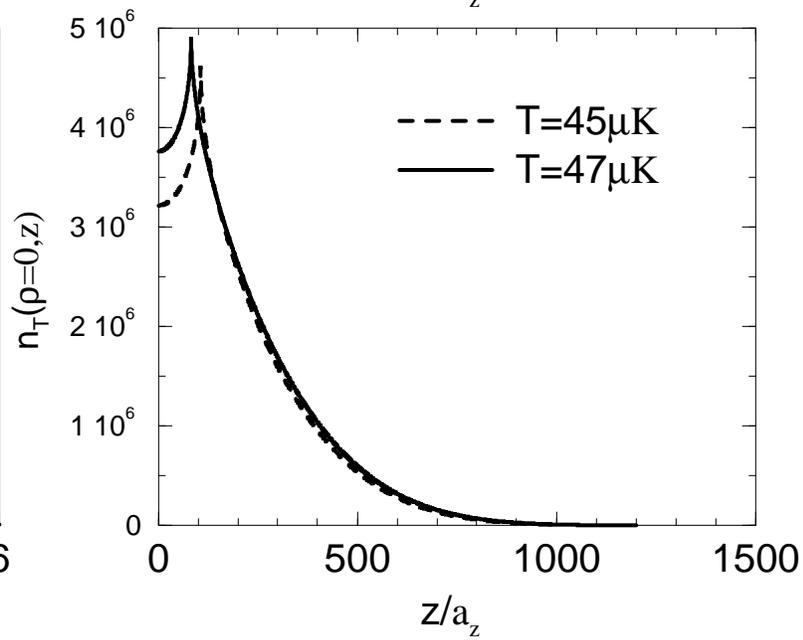
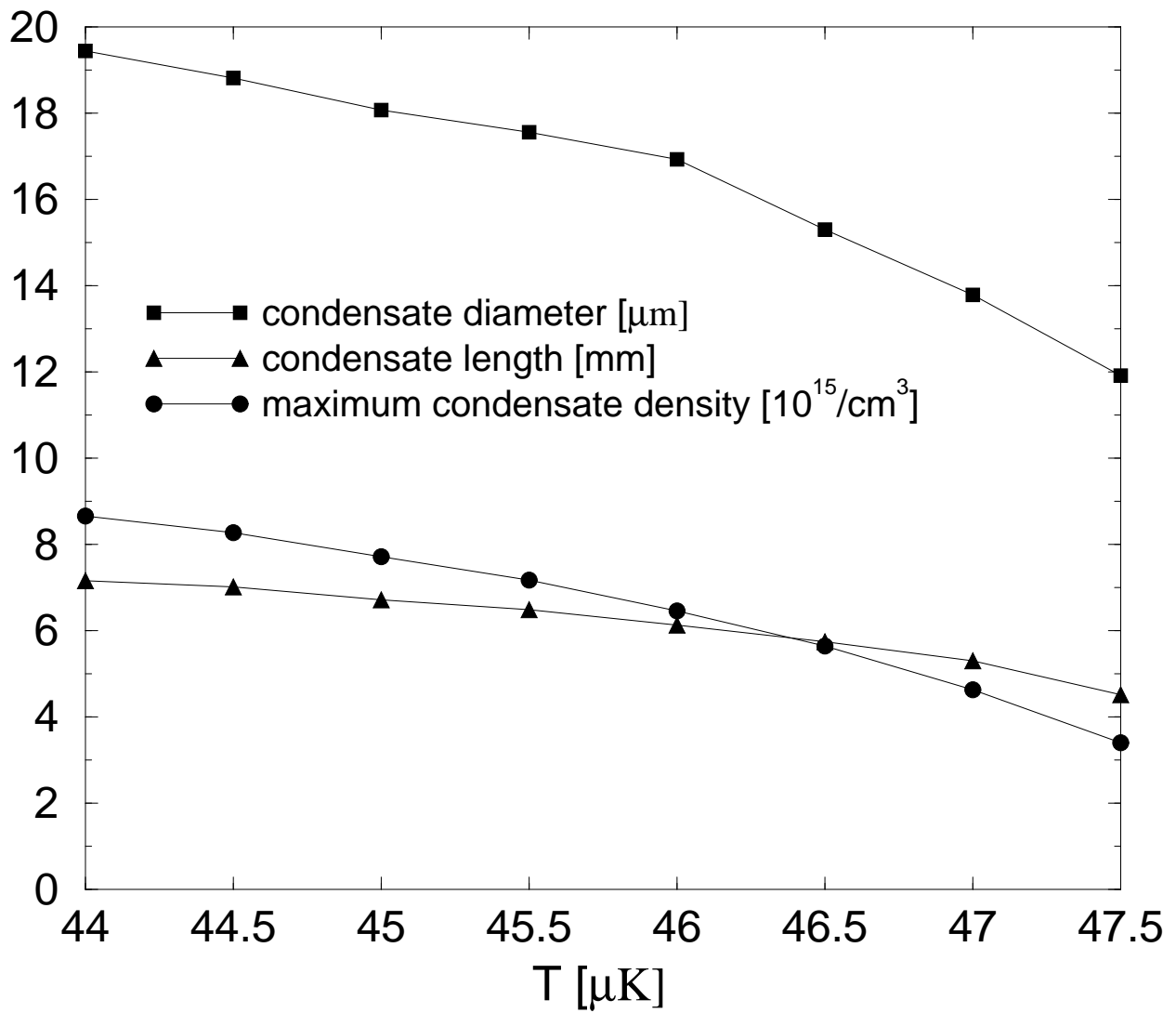
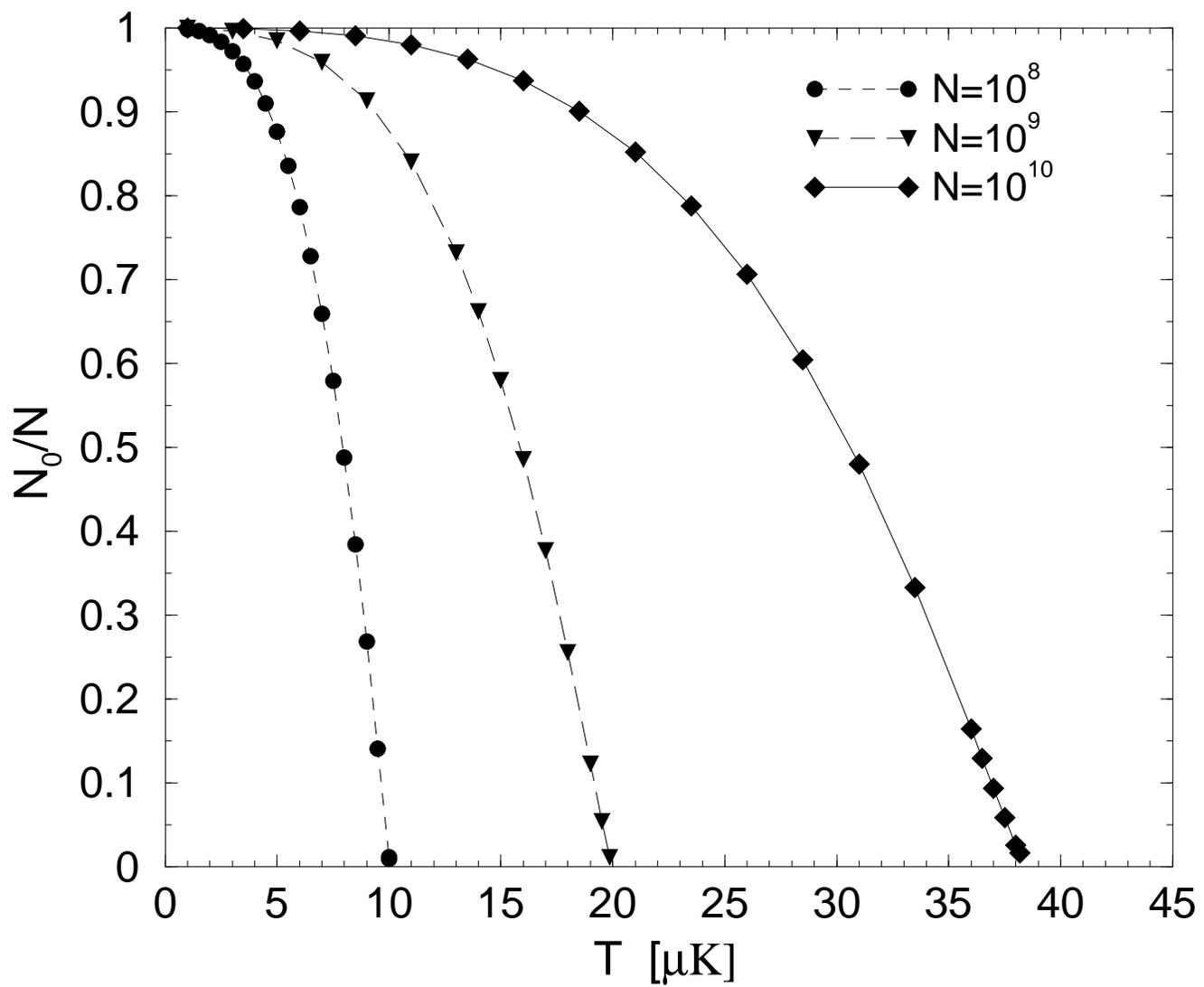
\end{document}